\documentclass[a4paper]{article}
\usepackage[a4paper,includemp=false,body={16cm,24.7cm},vcentering]{geometry}
\usepackage[latin1]{inputenc}
\usepackage[T1]{fontenc}
\usepackage{mathptmx} 
\usepackage[bf,center]{caption}
\usepackage{graphicx} 

\usepackage{url,amsmath,amssymb,amsthm}

\makeatletter
\renewcommand{\section}{\@startsection{section}{1}{0mm}{30pt}{12pt}{\normalfont\normalsize\bfseries}}

\renewcommand{\subsection}{\@startsection{subsection}{2}{0mm}{18pt}{12pt}{\normalfont\normalsize\itshape}}
\makeatother

\newcommand{\Title}[1]{\begin{center}{\bfseries\fontsize{12pt}{12pt}\selectfont#1}\end{center}}
\newcommand{\Author}[2]{\begin{center}{\fontsize{12pt}{12pt}\selectfont#1}\\{\it #2~}\end{center}}
\newcommand{\Introduction}{\section*{Introduction}}
\newcommand{\Conclusion}{\section*{Conclusion}}

\begin{document}

\Title{Stellar wobble in triple star systems}
  
\Author{M.H.M. Morais$^1$, A.C.M. Correia$^{1,2}$}{1. Department of Physics, I3N, University of Aveiro, \\  Campus Universitario de Santiago, 3810-193 Aveiro, Portugal,  helena.morais@ua.pt \\
2. Astronomie et Systemes Dynamiques, IMCCE-CNRS UMR 8028, \\  77 Avenue Denfert-Rochereau, 75014 Paris, France, correia@ua.pt  }
 
\Introduction

\noindent

The radial velocity method for detecting extra-solar planets relies on measuring the star's wobble around the system's center of mass. Since this is an indirect method, we may ask if there are other dynamical effects that can mimic such wobble. In recent articles\cite{1,2,3}, we modeled the effect of a nearby binary system on a star's radial velocity. We showed that, if we are unaware of this nearby binary, for instance because one component is unresolved or both components are faint stars, the binary's effect may mimic a planet. Here, we review this work, explaining in which circumstances the binary's effect may mimic a planet and we discuss what can be done in practice in order to distinguish between these two scenarios (planet or nearby binary). 

\section{Overview of theory and results}

\noindent

We study a triple system composed  by a star with mass $m_2$, at distance $\vec{r}_2$ from the centre of mass of a  binary with masses $m_{0}+m_{1}$, and inter-binary distance $\vec{r}_1$.  We assume that $\rho=|\vec{r_{1}}|/|\vec{r_{2}}|\ll 1$ (hierarchical system). 
The Hamiltonian in Jacobi coordinates is  \cite{3}
\begin{equation}
\label{hamilton}
H = -\frac{G m_{0} m_{1}}{2\,a_1} -\frac{G (m_{0}+m_{1}) m_{2}}{2\,a_2}+ F \ ,
\end{equation}
where an approximation of the perturbation term is
\begin{eqnarray}
\label{quadrupole}
F &=&-\frac{G m_2}{r_{2}} \frac{m_0 m_1}{m_0+m_1} \frac{\rho^2}{2} \left( 3\,\cos^2{S}-1 \right)  \ ,
\end{eqnarray}
with $S$ the angle between $\vec{r}_1$ and $\vec{r}_2$.
 
From Eq.~(\ref{hamilton})  we see that the motion is, approximately, a composition of two Keplerian orbits described by: 
$\vec{r}_1$, with semi-major axis $a_1$, eccentricity $e_1$, and period $T_1=2\,\pi/n_1$;  and $\vec{r}_2$, with semi-major axis $a_1$, eccentricity $e_1$, and period $T_1=2\,\pi/n_1$. 
 
The radial velocity of the star $m_2$ is  $V_{R}=V_{RK}+V_{RP}$, where $V_{RK}$  is a Keplerian term that describes the motion around a "star", of mass $m_0+m_1$, located at the binary system's centre of mass, and  $V_{RP}$ is a small perturbation\cite{1,2}. 

The radial velocity data of  the star $m_2$ is first  fitted with a Keplerian radial velocity curve, $V_{RK}$.
After subtracting $V_{RK}$, we are left with the perturbation term, $V_{RP}$, from which we could in principle infer the presence of the nearby binary. However, as we will show next, this is not always possible in practice.  

\subsection{Coplanar circular orbits}

\noindent

In the case of coplanar circular orbits we have\cite{1}
\begin{eqnarray}
V_{RP} & = & K_0 \cos(n_2\,t+\theta_{0}) 
+ K_1\, \cos((2\,n_{1}-3\,n_2)\,t+\theta_{1}) 
+ K_2\, \cos((2\,n_{1}-n_2)\,t+\theta_{2}) \ .
\end{eqnarray}
The term with frequency $n_2$ is  incorporated in the main Keplerian curve, $V_{RK}$.
Since $n_1\gg n_2$, the  2nd and 3rd terms have very close frequencies that can only be resolved if the observation timespan $t_{obs}\ge T_{2}/2$. If there is enough resolution and precision, we identify  both signals and we conclude that they should not be caused by planets (as such close orbits would be unstable). However, since $|K_1|=5\,|K_2|$, in practice, due to limited precision, we may only be able to observe the signal with frequency $2\,n_1-3\,n_2$. In this case, we may think there is a planet companion to the observed star.  

We simulated a coplanar triple system composed of $m_2=M_{\odot}$ on a circular orbit of period $T_2=22$~y, around  a binary $m_0=0.7\,M_{\odot}$ and $m_1=0.35\,M_{\odot}$ with a circular orbit of period  $T_1=411$~d. We computed radial velocity data points over $t_{obs}=11$~y at precision 0.8~m/s. In Fig.~1 we see a  signal   with frequency $2\,n_1-3\,n_2$ (amplitude 0.9~m/s at 223~d) that mimics a planet of $18\,M_E$.

\begin{figure}
\centering
\includegraphics[width=6cm,angle=-90]{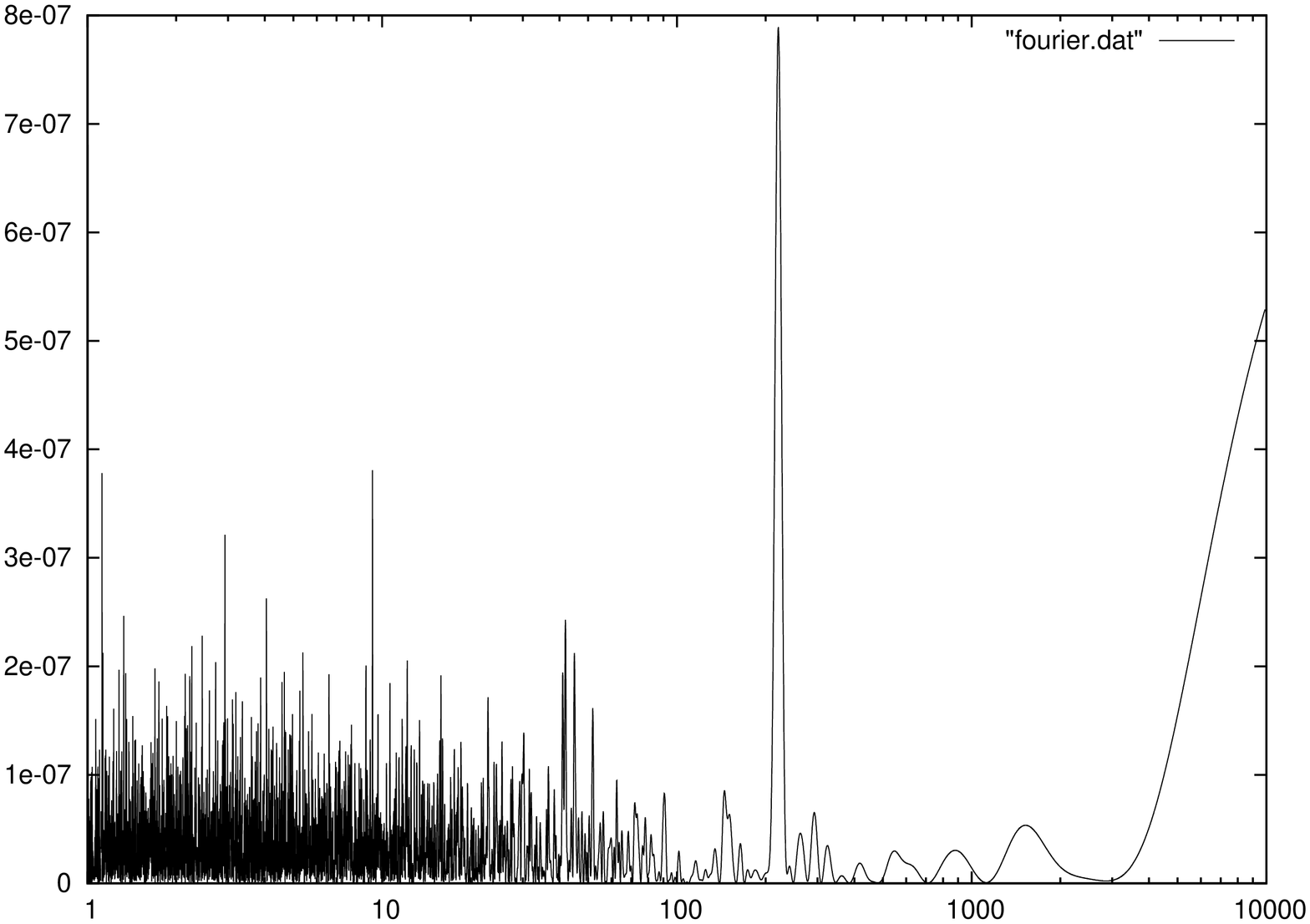}
\caption{Periodogram of residuals leftover after removing $V_{RK}$.}
\end{figure}

\subsection{Non-coplanar circular orbits}

\noindent

In the case of non-coplanar circular orbits, $V_{RP}$ is a combination of 6 periodic terms\cite{2} with frequencies: $n_1$, $3\,n_1$,  $2\,n_{1}\pm n_2$ and $2\,n_{1}\pm 3\,n_2$.
Depending on the observation's precision and resolution we may observe one or more of these terms. If all these have well separated frequencies we mistake them by planet(s). 

We simulated a triple system composed of $m_2=M_{\odot}$ on a circular orbit  of period $T_2=4.2$~y, around  a binary $m_0=m_1=0.25\,M_{\odot}$ with a circular orbit of period  $T_1=85$~d. The relative inclination is $i=30^\circ$. We computed radial velocity data points over $t_{obs}=11$~y at precision 
0.7~m/s. In Fig.~2 we see signals with frequency $2\,n_1-3\,n_2$ (amplitude 0.8~m/s at 46~d), and frequency $3\,n_2$ (amplitude 1.4~m/s at 516~d) that mimic planets of $7\,M_E$ and $20\,M_E$. 

\begin{figure}
\centering
\includegraphics[width=6cm,angle=-90]{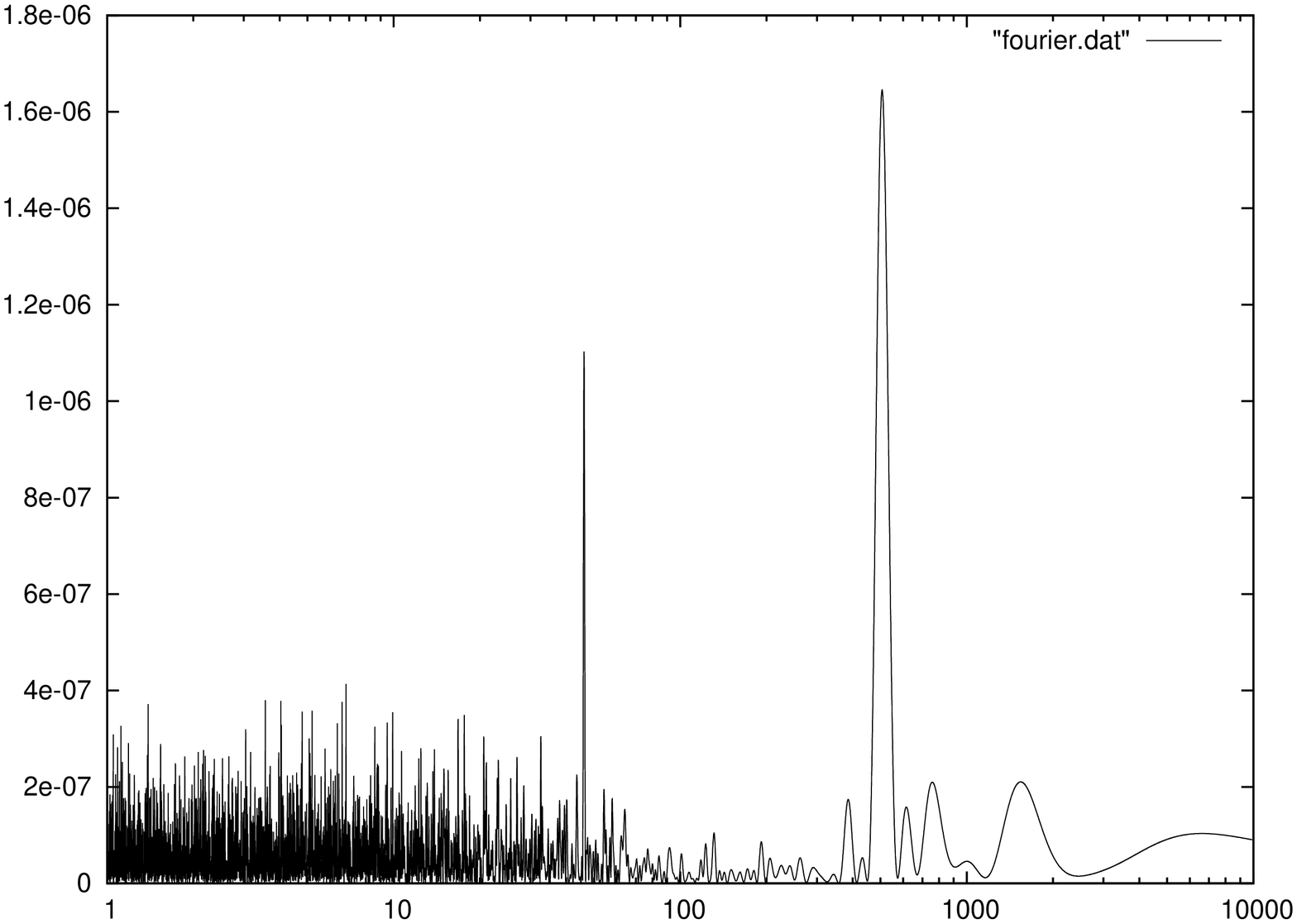}
\caption{Periodogram of residuals leftover after removing $V_{RK}$.}
\end{figure}

\subsection{Eccentric coplanar orbits}

\noindent

It can be shown that, generally, $V_{RP}$ is  a composition of short period terms\cite{2}. When the orbits are circular we saw above that there are finite number of periodic terms. When the orbits are eccentric we must express 
$V_{RP}$ as an expansion in $e_1$ and $e_2$, hence  there is an infinite number of periodic terms. However, in practice only a finite number of periodic terms (the ones that appear at lowest order in $e_1$ and $e_2$) are important.  
In the eccentric 2D case, up to 1st order in the eccentricities, there are 12 frequencies\cite{2}: $n_2$, $2\,n_{1}-n_2$, $2\,n_{1}-3\,n_2$, $n_{1}-n_2$, $n_{1}-3\,n_2$, $3\,n_{1}-n_2$, $3\,n_{1}-3\,n_2$, $n_{1}+n_2$, $2\,n_2$, $2\,n_{1}$, $2\,n_{1}-4\,n_2$ and $2\,n_{1}-2\,n_2$.

We simulated a coplanar triple system composed of a star $m_{2}=M_{\odot}$ with $e_2=0.1$ and period $T_2=22$~y, in orbit around a binary with $m_0=0.7\,M_{\odot}$ and $m_1=0.35\,M_{\odot}$, $e_1=0.2$ and period  $T_1=411$~d.  We computed radial velocity data points over $t_{obs}=11$~y at precision 0.8~m/s.
In Fig.~3 we see  signals with frequencies $2\,n_1-3\,n_2$ (amplitude 0.8~m/s at 223~d) and $n_1-3\,n_2$ (amplitude 1.4~m/s at 487~d). Since $n_2 \ll n_1$ these mimic  planets of $15\,M_E$ and $34\,M_E$ at the 2/1 mean motion resonance. 

\begin{figure}
\centering
\includegraphics[width=6cm,angle=-90]{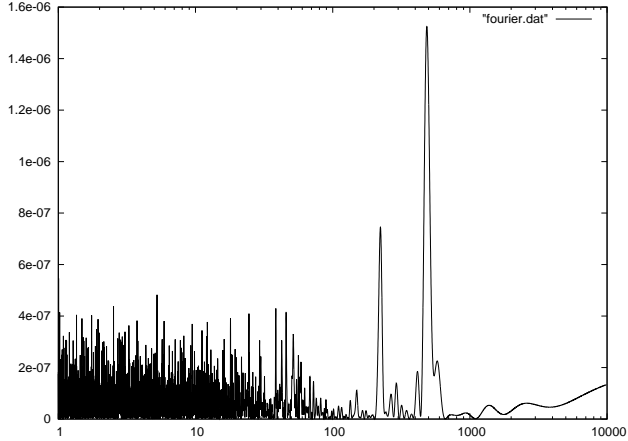}
\caption{Periodogram of residuals leftover after removing $V_{RK}$.}
\end{figure}

\subsection{Secular evolution}

If the star $m_2$ has an eccentric orbit ($e_2\neq 0$) and if $t_{obs}>T_2$, signals at or nearby harmonics of  $n_2$ appear\cite{2,3}. These may be mistaken by planet(s) near a mean motion resonance with a companion "star" of mass $m_0+m_1$\cite{3}.  However, as $t_{obs}$  increases, the short period terms should become negligible with respect to the orbits' secular evolution. 

The secular evolution can be obtained by averaging Eq.~(\ref{hamilton}) with respect to the orbital periods $T_1$ and $T_2$. Moreover, since $a_1\ll a_2$ (hierarchical system) and $m_2>m_0+m_1$ (perturbing binary less massive than observed star), it can be shown that the star's motion around the binary's center of mass coincides, approximately, with the invariant plane\cite{3}. The secular Hamiltonian, obtaining after averaging equation (2) over the mean motion of both orbits, is approximately\footnote{In \cite{3} we present a longer but more correct derivation of the triple system's secular evolution.}
\begin{equation}
\label{hamiltonian0}
F_{sec}= \frac{G}{16}  \frac{m_0 m_1}{m_0+m_1}\frac{m_2}{(1-e_2^2)^{3/2}} \frac{a_1^2}{a_2^3} 
\left[ (2+3\,e_{1}^2)(3\,\cos^2{i}-1)+15\,e_{1}^2\,\sin^2{i}\,\cos(2\,\omega_1) \right] 
\end{equation}
with $e_1$ (binary's eccentricity), $\omega_1$ (binary's argument of pericentre), $i$ (relative inclination). 

From Hamilton's equations we obtain the secular evolution of the star's orbit.
Since Eq.~(\ref{hamiltonian0}) does not depend on $\omega_2$ (star's  argument of pericentre) then $\dot{e}_2=0$ and 
\begin{equation}
\label{precession}
\dot{\omega}_2 = \frac{3}{4}\,A\,\frac{m_0\,m_1}{(m_0+m_1)^2}\,\left( \frac{a_1}{a_2} \right)^2 \frac{n_2}{(1-e_2^2)^2} \ ,
\end{equation}
where
\begin{equation}
\label{factor}
 A=  \left( \frac{1}{2}+\frac{3}{4}\,e_{1}^2 \right)(3\,\cos^2{i}-1)+\frac{15}{4}\,e_{1}^2(1-\cos^2{i})\cos(2\,\omega_1) 
\end{equation}
depends on the binary's ($m_0+m_1$) orbit. 

In Fig.~4 we show  $A$ calculated at fixed $e_1$\cite{3}. If $i<40^\circ$ we can eliminate $\omega_1$ from Eq.~(\ref{factor}) since $\omega_1$ circulates. 
If $i>40^\circ$ we place the binary at the Kozai equilibrium ($e_{1}=\sqrt{1-(5/3)\,\cos^2{i}}$, $\omega_1=\pm90^\circ$).
\begin{figure}
  \centering
    \includegraphics[width=9cm]{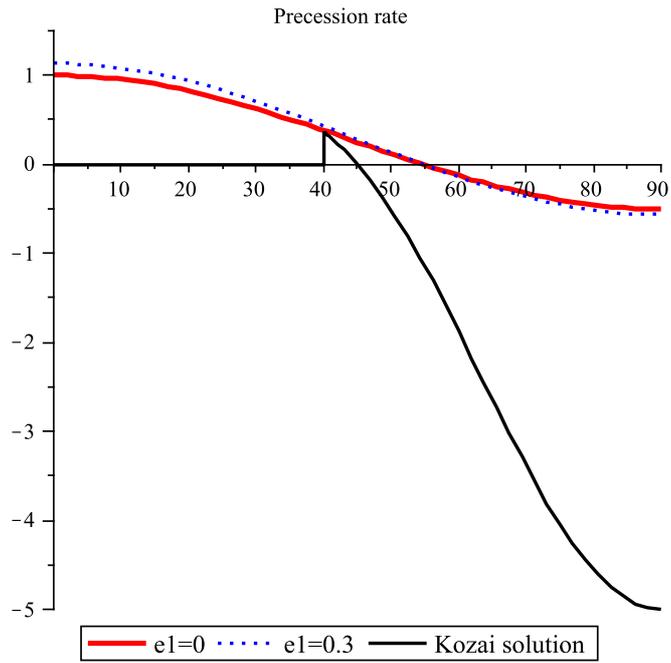}
  \caption{The normalized precession rate ($A$).}
 \end{figure}

Therefore,  $V_R$ is better fitted by the radial velocity curve of a precessing ellipse\cite{3}
\begin{equation}
\label{vr}
V_R =  \frac{n_2\,a_2}{(1-e_2^2)^{1/2}} \frac{m_b\,\sin{I_2}}{m_2+m_b}
 \left( \cos(f_2+\omega_{2,0}+\dot \omega_2 t) + e_2\,\cos(\omega_{2,0}+\dot \omega_2 t) \right)
\end{equation}
where $m_b=m_0+m_1$ and $\dot \omega_2$ is given by Eq.~(\ref{precession}). 

We simulated a system composed of a star $m_2=M_{\odot}$ with an eccentric orbit ($e_2=0.2$) of period $T_2=4.24$~y, and a binary with masses $m_0=0.35\,M_{\odot}$ and $m_1=0.15\,M_{\odot}$,  period $T_1=85$~d, relative inclination  $i=60^{\circ}$ and  eccentricity $e_1=0.76$ (at the Kozai equilibrium).  We computed radial velocity data points over $t_{obs}=6$~y at precision 5~m/s.
In Fig.~5 (top) we see  the residuals leftover after fitting a fixed Keplerian radial velocity curve. There is an an obvious peak at 606~day which is a frequency close to an harmonic of $n_2$. In Fig.~5 (low) we see  the residuals leftover after fitting a radial velocity curve with precession (Eqs.~\ref{vr},\ref{precession}). We obtain $\dot{\omega}_2=-0.25^\circ$/y and the peak  at 606~d  is no longer important.

\begin{figure}
  \centering
    \includegraphics[width=9cm]{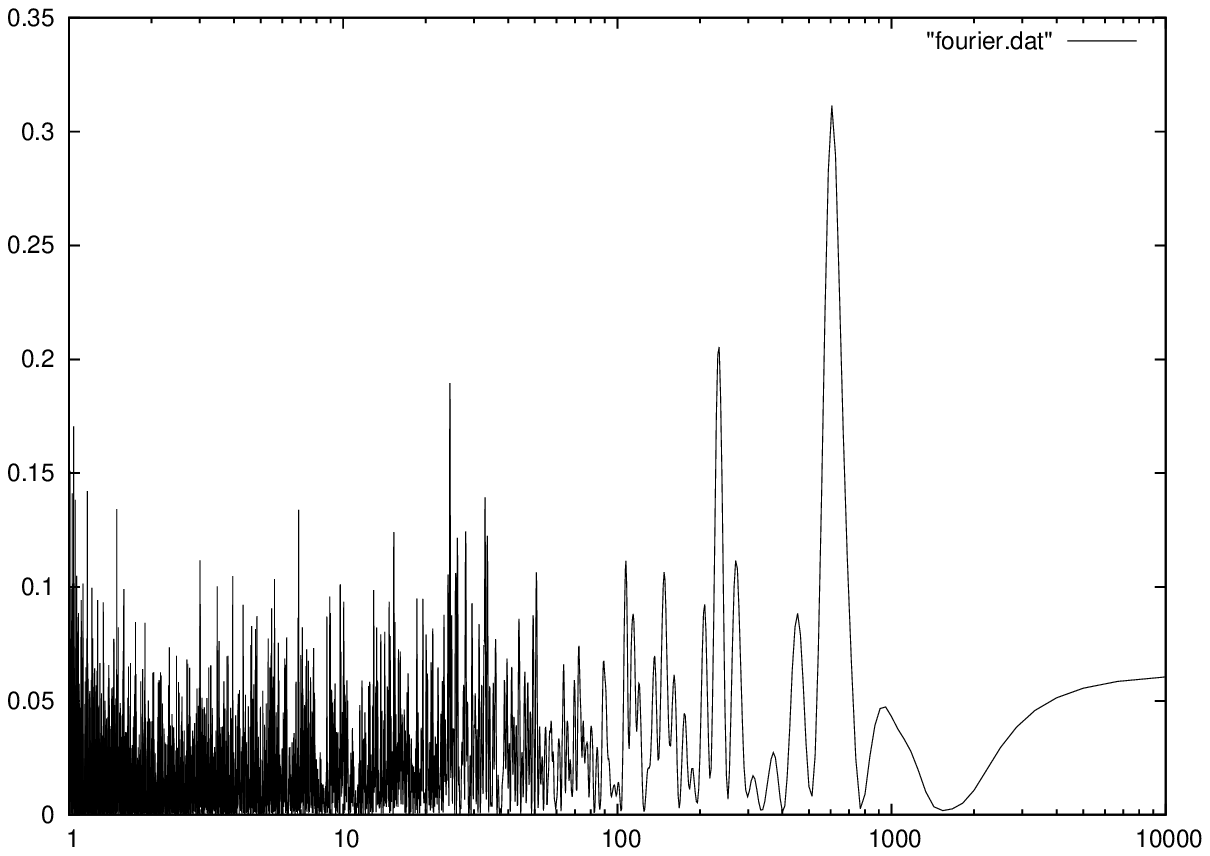}
    \includegraphics[width=9cm]{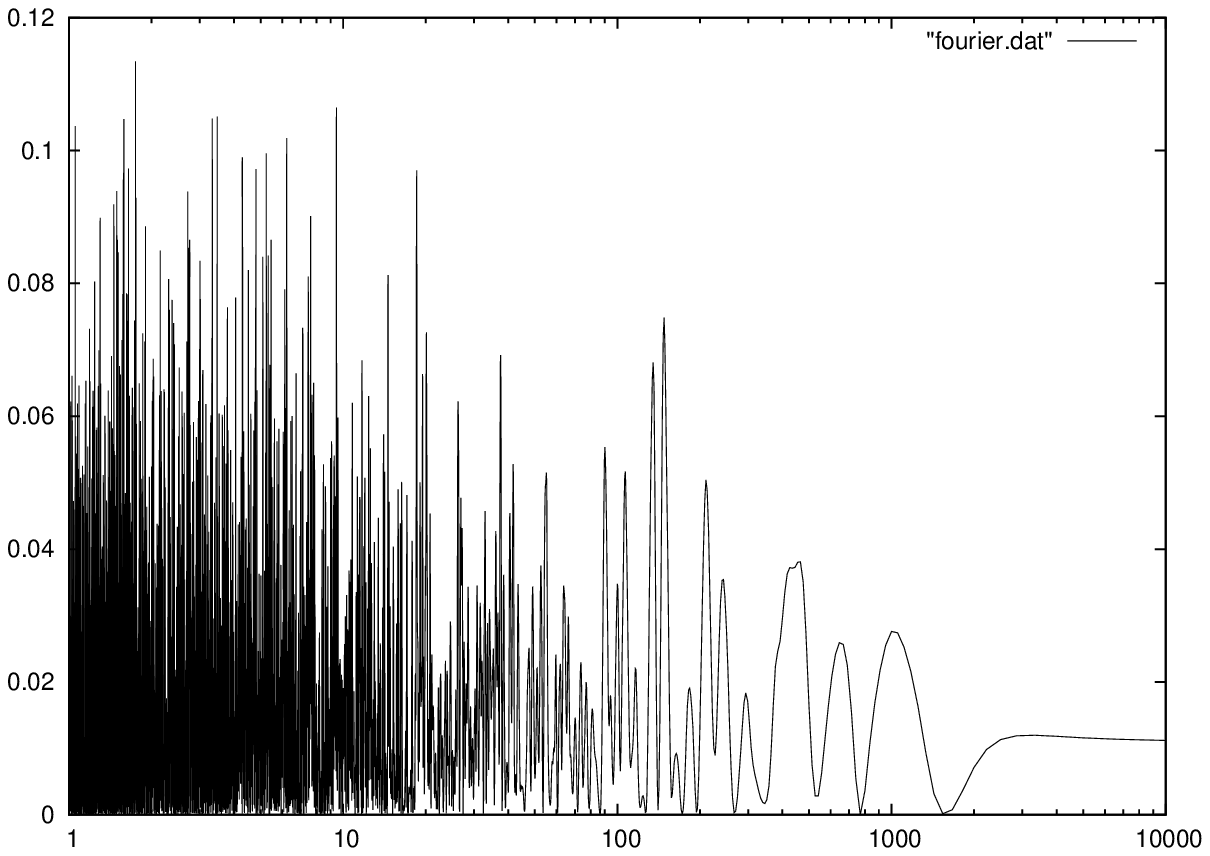} 
  \caption{Periodogram of residuals leftover after fitting Keplerian radial velocity curve (top) or radial velocity curve of precessing ellipse (low) to the data.}
  \end{figure}

\Conclusion

\noindent

When $t_{obs}<T_2$, we observe short-period terms due to the binary that may be mistaken by planet(s).
In order to distinguish the binary's effect from planet(s) we need precise observations over a reasonable long timespan, which is often not possible in practice.
However, a signal with frequency, $n_{pl}$, and amplitude, $K_{pl}$, mimics a planet with parameters\cite{1,2}
\begin{eqnarray}
a_{pl} & =& \frac{(G \,m_{2})^{1/3}}{n_{pl}^{2/3}} \ , \\ 
 m_{pl}\,\sin{I_{pl}} &=& K_{pl}\,\frac{m_{0}+m_1+m_2}{n_{pl}\,a_{pl}} \ .
\end{eqnarray}
Therefore, when the detected signals' frequencies are well separated
we can invert these expressions to predict binary system's parameters and check if they are realistic\cite{1,2}.

When $t_{obs}\gg T_2$ the orbits' secular evolution dominates over the short-period terms. In this case, we showed that the star's orbit precesses due to the binary, and measuring this precession rate provides estimates for the binary's parameters\cite{3}. 

We detected such a precessional effect in the radial velocity data of the star $\nu$-Octantis~A\cite{3}. This could provide an alternative explanation for the signal at 417~d that had previously been identified as a planet\cite{4}. Since $\nu$-Octantis~A  has a nearby companion ($\nu$-Octantis~B) on a $2.8$~y orbit\cite{4}, a planet at 417~d cannot be stable on a coplanar prograde orbit (although it can survive on a coplanar retrograde orbit)\cite{5}.  We measured retrograde precession of $-0.86^\circ$/y which could be explained if $\nu$-0ctantis~B was a double star inclined $i>45^\circ$ with respect to the main binary's orbit\cite{3}. Moreover, we observed that after fitting a precessing orbit to the radial velocity data, the signal at 417~d was no longer prominent in the leftover residuals\cite{3}. Although, retrograde precession of the main binary could also be explained by a planet on a highly inclined orbit\cite{3}, these orbits do not seem to be stable\cite{3}.


\begin{thebibliography}{99}

\bibitem{1} Morais, M.H.M., A.C.M. Correia, 2008, A\&A, Vol 391, 899

\bibitem{2} Morais, M.H.M., A.C.M. Correia, 2011, A\&A, Vol 525, A152

\bibitem{3} Morais, M.H.M., A.C.M. Correia, 2012, MNRAS, Vol 419, 3447.

\bibitem{4} Ramm, D.J.  et al 2009, MNRAS, Vol 394, 1695

\bibitem{5} Quarles, B., Z.E. Musielak, M. Cuntz, 2012, MNRAS, in press.


\end{thebibliography}
\end{document}